\documentclass[twocolumn,floatfix,superscriptaddress,longbibliography,PRL]{revtex4}
\usepackage{amsfonts,amssymb,amsmath,hyperref}
\usepackage[applemac]{inputenc}
\usepackage{multirow}
\usepackage{color}
\usepackage{dcolumn}
\usepackage{bm}
\usepackage{array}
\usepackage{tabularx}
\newcolumntype{Y}{>{\centering\arraybackslash}X}

\usepackage[paperwidth=210mm,paperheight=297mm,centering,hmargin=2cm,vmargin=2cm]{geometry}

\newif\ifhyper
\hypertrue
\ifhyper
\hypersetup{
citecolor = {green},
colorlinks = {true}, 
urlcolor = {blue} 
}
\fi

\newlength{\ldag}
\def\be{\begin{equation}}
\def\ee{\end{equation}}
\def\bea{\begin{eqnarray}}
\def\eea{\end{eqnarray}}
\def\bse{\begin{subequations}}
\def\ese{\end{subequations}}
\def\bc{\begin{center}}
\def\ec{\end{center}}

\allowdisplaybreaks 
\begin{document}

\title{Auxiliary fields approach to shift-symmetric  theories: The  $\bm \varphi^4$ derivative theory  and  the crumpled-to-flat transition of membranes at two-loop order}

\author{L. Delzescaux} 
\email{louise.delzescaux@sorbonne-universite.fr}
\affiliation{Sorbonne Universit\'e, CNRS, Laboratoire de Physique Th\'eorique de la Mati\`ere Condens\'ee, LPTMC, 75005 Paris, France}

\author{C. Duclut} 
\email{charlie.duclut@curie.fr}
\affiliation{Laboratoire Physico-Chimie Curie, CNRS UMR 168, Institut Curie,
Universit\'e PSL, Sorbonne Universit\'e, 75005, Paris, France}

\author{D. Mouhanna} 
\email{mouhanna@lptmc.jussieu.fr}
\affiliation{Sorbonne Universit\'e, CNRS, Laboratoire de Physique Th\'eorique de la Mati\`ere Condens\'ee, LPTMC, 75005 Paris, France}

\author{M. Tissier} 
\email{tissier@lptmc.jussieu.fr}
\affiliation{Sorbonne Universit\'e, CNRS, Laboratoire de Physique Th\'eorique de la Mati\`ere Condens\'ee, LPTMC, 75005 Paris, France}


\begin{abstract}

We introduce a technique relying  on the use of auxiliary  fields in order to eliminate  explicit field-derivatives  that plague the high orders renormalization group treatment of shift-symmetric (derivative) theories. This  technique   simplifies drastically the computation of fluctuations in such theories. This  is illustrated  by  deriving   the two-loop renormalization group equations -- and the three-loop anomalous dimension --  of the $\bm\varphi^4$ derivative  theory  in $D=4-\epsilon$,  which is also relevant to describe the crumpled-to-flat  transition  of polymerized membranes.  Some features of this transition are provided.

\end{abstract}

\maketitle 

\begin{center} 
{\bf I. INTRODUCTION}
\end{center}

Shift-symmetric theories or  shift-symmetric   enforced preexisting theories  are characterized, in the simplest situation,  by  an invariance of the kind 
\begin{equation}
\bm\varphi\longrightarrow \bm \varphi + \bf c
\end{equation}
where  $\bf c$  is a constant vector. Such theories  have been the subject of a strong activity in the past and have received a renewed interest in recent years,  mainly in the context  of the study of modified theories of gravity, as well as in string and brane theories. A strong motivation is that shift symmetry  results in specific renormalization properties. This is for instance  the case of  Galileon field theory \cite{nicolis09}  that displays  field and space-time  shift symmetries  $\varphi\longrightarrow \varphi + c + a_{\mu} x_{\mu}$ where $c$ and the $a_{\mu}$ are constants. It obeys  nonrenormalization theorems \cite{luty03,hinterbichler10}; see  \cite{Goon16} for a  review  and  \cite{codello13} for a nonperturbative treatment. In  the context of Hor$\check{\hbox{a}}$va-Lifshitz gravity \cite{horava09} it has been shown that  shift symmetry prevents the appearance of an infinite number of interactions \cite{fujimori15}.  Generalized -- polynomial -- shift symmetry \cite{griffin13,hinterbichler14,griffin15}  has also been considered, notably in connection with multicritical symmetry breaking. Very recently,  shift symmetry has been considered  in the context of asymptotically safe quantum gravity-matter theories where this symmetry allows generating closed renormalization group (RG) flow for the effective action, see e.g.\cite{laporte21}. Interestingly, motivated by these considerations,  a new universality class has been discovered that gathers theories whose RG equations are projected on functions of the kinetic term \cite{laporte23}. Shift symmetry have also been studied in various other contexts including  Horndeski gravity~\cite{traykova21,eichhorn23},  inflation~\cite{brax05,finelli18}, inflation in supergravity~\cite{brax05} and in anti-de Sitter space~\cite{bonifacio19}.

\begin{center} 
{\bf II. MEMBRANES }
\end{center}

Shift symmetry is also relevant  in condensed matter physics through the long distance description of  both fluid and polymerized membranes, see \cite{proceedings89,bowick01} for reviews. In particular, polymerized membranes have been intensively investigated  these last twenty years following the discovery of graphene~\cite{novoselov04,novoselov05} and graphene-like materials, see e.g.~\cite{katsnelson12}. For a $D$-dimensional membrane embedded in a $d$-dimensional Euclidean space, the parametrization of a point ${\bf x}\in \mathbb{R}^D$ in the  membrane is realized through the mapping ${\bf x}\to {\bf R} ({\bf x})$ where  ${\bf R}$ is  a field in $\mathbb{R}^d$. For obvious reasons,  the energy of the membrane can only depend on variations  of ${\bf R}$ so that   the action  should display  a shift symmetry ${\bf R}(\bf x)\to {\bf R}(\bf x)+{\bf C}$ where ${\bf C}$ is a constant vector.   The  relevant action to study  polymerized membranes in $D$ dimensions is given by~\cite{paczuski88}:
\begin{align}
\begin{split}
S[\{\bm R\}] = \int & \text{d}^Dx \ \bigg\{ \frac{\kappa}{2} ({\partial_\alpha^2}{\bf R})^2 + \frac{{r}}{2} ({\partial_\alpha}{\bf R})^2 +  \\ 
&  \hspace{-0.5cm}+\frac{\lambda}{8} (\partial_\alpha \bm{R}.\partial_\alpha{\bf R})^2 + \frac{\mu}{4}\,(\partial_\alpha {\bf R}.\partial_\beta{\bf R})^2 \bigg\} 
\label{S}    
\end{split}
\end{align}
where Greek indices run over $1\dots D$ and summation over repeated indices is implicit.  In  Eq.~(\ref{S}), $\kappa$ is the bending rigidity constant, $r$ is a tension coefficient that conveys the main temperature dependence. The coefficients $\lambda$ and $\mu$, which are associated with quartic interactions,  are  Lam\'e (elasticity) coefficients that embody elasticity and shear properties of the membrane; stability requires $\kappa$, $\mu$, and the bulk modulus $B=\lambda+2 \mu/D$ to all be positive. In agreement with shift symmetry,   action (\ref{S})  is expressed purely in terms of field-derivatives.  It displays an invariance under the action of  the Euclidean group of  displacements E($d$)   that includes  both rotations  and translations  in $d$ dimensions: $R_{\mu}\longrightarrow {\cal R}_{\mu \nu} R_{\nu} + C_{\mu}$, where ${\cal R}$ is a rotation matrix and  $\bf C$ a  constant  vector,  see e.g.~\cite{zanusso14,coquand19b}.  The model  (\ref{S}) is directly relevant to study the crumpled-to-flat transition in polymerized membranes, see  \cite{paczuski88}.  Indeed, varying the tension coefficient $r$, one expects a phase transition between a disordered, crumpled, phase at high temperatures and an ordered, flat, phase at low temperatures, characterized by a well-defined orientation of the membrane and, thus, a nonvanishing average value of the tangent vector fields $t_{\alpha} = \partial_{\alpha}{\bf R}$. This transition should occur in any dimension $D$ higher than the lower critical dimension $D_{lc}(d)$ that has been evaluated in \cite{kownacki09} for an Euclidean embedding space of dimensions $d = 3$ to be close to 1.33 \footnote{The fact that $D_{lc}(d = 3)$ is lower than 2 relies on the long-range character of the interactions that are induced by their derivative character. This allows to bypass the Hohenberg-Mermin-Wagner theorem and to find an ordered phase even in $D = 2$.}. Note however that, in a realistic model of membranes, self-avoidance very likely destroys the crumpled phase, see e.g.~\cite{gompper97}, so that membranes always lie in their flat phase. The properties of the crumpled-to-flat transition have been studied at one-loop in the vicinity of the upper critical dimension $D=4$ in \cite{paczuski88}, within  a large-$d$ expansion in \cite{david88,aronovitz89}, by self-consistent screening approximation  (SCSA) \cite{ledoussal92,ledoussal18} as well as within a nonperturbative framework in \cite{kownacki09,braghin10,hasselmann11,essafi11,essafi14,coquand18}. Note finally  that, prior to its role as long distance effective action for membranes,  action (\ref{S}) is nothing but the derivative version of the $O(d)$ $\bm\varphi^4$-model and, as such, represents the simplest but nontrivial model of shift-symmetric derivative field theory. 

Starting from  action (\ref{S}), it is also possible to  study the flat phase of membranes \cite{nelson87,aronovitz88,david88,aronovitz89,guitter89,gornyi15,ledoussal92,gazit09,zakharchenko10,roldan11,ledoussal18} which is  relevant to investigate the properties of stable  graphene and graphene-like materials as well as that of biological membranes endowed with a cytoskeleton. To do that, one considers a flat configuration given by ${\bf R}^0=({\bf x}, {\bf 0}_{d_c})$ where ${\bf 0}_{d_c}$ is the null vector of  dimension $d_c=d-D$,  and  decomposes the field  $\bf R$ into $\bf R(x)=[\bf x+{\bf u}(\bf x), {\bf h}(\bf x)]$ where ${\bf u}$ and ${\bf h}$ represent $D$ longitudinal, phonon, modes  and  $d-D$ transverse, flexural,  modes, respectively. Power-counting considerations then lead to the relevant action written in terms of phonon and flexural modes \cite{nelson87,aronovitz88,david88,aronovitz89,guitter89,gornyi15,ledoussal92,gazit09,zakharchenko10,roldan11,ledoussal18}: 
\begin{equation}
\begin{array}{ll}
\hspace{-0.3cm} S[{\bf h,u}]=\displaystyle \int \text{d}^Dx  & \hspace{-0.1cm} \displaystyle \left\{ {\kappa \over 2}\big(\partial_\alpha^2 {\bf h} \big)^2 
+ {\lambda\over 2}\, u_{\alpha\alpha}^2 + {\mu}\, u_{\alpha\beta}^2 \right\},
\label{action}
\end{array}
\end{equation}
where $u_{\alpha\beta}$ is the strain tensor that encodes the elastic fluctuations around the flat phase configuration ${\bf R}^0({\bf x)}$:
$u_{\alpha\beta}={1\over 2}(\partial_{\alpha}{\bf R} . \partial_{\beta}{\bf R}-\partial_{\alpha}{\bf R}^0 . \partial_{\beta}{\bf R}^0)={1\over 2}(\partial_{\alpha} {\bf R} . \partial_{\beta}{\bf R}- \delta_{\alpha\beta})$. It is given by,  neglecting nonlinearities in the phonon field ${\bf u}$: 
\begin{equation}
u_{\alpha\beta} \simeq {1\over 2} \left[\partial_\alpha u_\beta+\partial_\beta u_\alpha+ \partial_\alpha {\bf h} . \partial_\beta {\bf h} \right]\ . 
\label{stress}
\end{equation}
Action (\ref{action}) has  been investigated perturbatively in the vicinity of  the upper critical dimension $D=4$  in~\cite{aronovitz88,aronovitz89,guitter89, coquand20a,mauri20,metayer22,pikelner22,metayer23};  see  \cite{coquand20b,metayer22d}  for investigation of the disordered case. It has also been studied by means of large-$d$ expansions \cite{david88,guitter88,aronovitz89,guitter89,gornyi15,saykin20}, SCSA \cite{ledoussal92,gazit09,ledoussal18} and within  a nonperturbative approach \cite{kownacki09,braghin10,hasselmann11}.

\begin{center} 
{\bf III. SCALE AND CONFORMAL INVARIANCE }
\end{center}

The two models~(\ref{S}) and~(\ref{action}) and, more generally, shift-symmetric theories have recently been the subject of a special attention regarding   the question of the relation between scale and conformal invariance, see~\cite{nakayama15} for a review. Indeed, it is generally believed that, at least in 2 and 4 dimensions,  conformal invariance  follows from scale invariance  for theories displaying unitarity --  in Euclidean space, reflection positivity -- and Poincar\'e  invariance~\cite{zamolodchikov86,polchinski88,jack90}. Also, for  theories that are suspected to be nonunitary -- which is the case for the actions~(\ref{S}) and~(\ref{action}) --   this property  no longer extends straightforwardly. For instance,  Riva and Cardy~\cite{riva05} have exhibited a model of elasticity --  whose action (\ref{action}) is a generalization embedded  in a Euclidean $d$-dimensional space -- which exhibits scale invariance but not conformal invariance.  One notes that Riva and Cardy have considered a -- nonunitary -- free theory, see also~\cite{elshowk11},  while one expects conformal invariance for  interacting  theories. Recently,  Safari  {\it et al.}~\cite{safari22} have investigated a large class of  derivative shift-symmetric,  scalar, nonunitary  theories, including the one described by action~(\ref{S}) for $d=1$; see also~\cite{tseytlin23}. By computing the trace of the momentum-energy tensor,  they have  shown that these theories were displaying  conformal invariance at the fixed point, in agreement with the common belief.  Conversely,  Mauri and Katsnelson~\cite{mauri21} have investigated the $\bm\varphi^4$ vectorial, derivative theory in its flat phase described by action~(\ref{action}).  On the same basis,  they have  concluded that  dilatation invariance  at the fixed point does not  extend to full conformal invariance at the fully attractive fixed point controlling the low temperatures, flat, phase. If this is confirmed, this would constitute an uncommon  example of interacting but non conformally invariant model, see~\cite{delzescaux23b}. 

\begin{center} 
{\bf IV. BEYOND ONE LOOP ORDER }
\end{center}

 These results, as well as almost all results derived in the context of shift-symmetric, derivative field theories,  have  been  obtained  at leading order \color{red}: \color{black} one-loop order in the perturbative context or  using  a low-derivative truncated  action   in the context   of the nonperturbative RG~\cite{dupuis21} \footnote{Within the nonperturbative RG approach there is generally no  recourse to expansion in powers of the usual parameters like  coupling constants, temperature, inverse of the number of components. However, the action is often expanded both in powers of the field(s) and derivatives of the field(s).}. There is a notable exception: the ${\bm\varphi}^4$ derivative theory in its ordered phase, described by action (\ref{action}), which has been investigated at high orders in perturbation theory. A characteristic  of this model  is  that its propagators  are given by \cite{aronovitz89}:
\begin{align}
\displaystyle & G_h^{ij}(p)=\langle h^{i}(p)h^{j}(-p)\rangle={\delta^{ij}\over \kappa p^4}  \\
& \displaystyle G_u^{\alpha\beta}(p)=\langle u^{\alpha}(p) u^{\beta}(-p)\rangle={P^{\perp}_{\alpha\beta}\over \mu p^2} +{P^{\parallel}_{\alpha\beta}\over (\lambda+2\mu) p^2} \nonumber
\end{align}
where $P^{\parallel}_{\alpha\beta}=p_{\alpha}p_{\beta}/ {\bm p}^2$,  $P^{\perp}_{\alpha\beta}=\delta_{\alpha\beta}-p_{\alpha}p_{\beta}/ {\bm p}^2$ while the indices $i,j$ run over $1\dots d-D$ and $\alpha, \beta$ over $1\dots D$. One sees on these expressions that $G_u^{\alpha\beta}(p)$ involves all  the coupling constants associated with  the interactions. Thus, considering the  two-point functions is  sufficient to get  the renormalization of all these couplings. This explains why this model has been investigated, although thirty years after  the one-loop computation  of Aronovitz and Lubensky \cite{aronovitz88},  successively  at two \cite{coquand20a,mauri20}, three \cite{metayer22,metayer23} and four loop order \cite{pikelner22}.

Clearly, it would be also very valuable to be able to extend the one-loop computation performed  by Paczuski and Kardar \cite{paczuski89} on the action (\ref{S})  describing the crumpled-to-flat transition at higher orders.  However, one faces here  an important difficulty:  the  derivative nature of  the interaction  makes the RG treatment  of action (\ref{S}) extremely  tedious if it is addressed by {\it brute force}. Indeed,  the propagator of the field $\bm R$ is,  in this case,
\begin{equation}
\begin{array}{ll}
\displaystyle G_R^{ij}(p)=\langle R^{i}(p)R^{j}(-p)\rangle={\delta^{ij}\over \kappa p^4}.\\
\end{array}
\end{equation}
It involves only the coupling $\kappa$ and to get the renormalization of the Lam\'e coefficients, one has to resort to four-point functions. Also,  the  derivative or, in Fourier space, momentum dependence  of the four-point vertices  gives rise to Feynman diagrams involving a very complex structure and a kinematic of the external momenta  that  are extremely  difficult to manage beyond one-loop order despite the use of   mathematical software like  {\it Mathematica}   and its package  LITERED \cite{Lee:2012cn,Lee:2013mka} to reduce the loop integrals. This is the reason why this model has been only studied at leading order in a loop-expansion. 

\begin{center} 
{\bf V.  AUXILIARY FIELDS  TECHNIQUE}
\end{center}

 We propose here a novel treatment  of derivative shift-symmetric theories that makes  their RG treatment  as easy  as that  of their nonderivative counterparts.   The main -- and simple -- idea is  that a derivative theory can be completely  reparametrized in terms of   auxiliary fields that represent the space-derivative of the original field(s). We basically follow the procedure first discussed by Faddeev and Popov in the framework of gauge theories \cite{Faddeev:1967fc}, later applied in disordered systems \cite{parisi79} or dynamical theories \cite{PhysRevA.8.423,Janssen76,dedominicis76}. Thanks to this trick, we are brought back to a theory where there are no more derivative interactions.   We apply this technique to the $\bm\varphi^4$ derivative theory  (\ref{S}) and derive the RG equations for the two coupling constants and tension coefficient  at two-loop order as well as the anomalous dimension at  the first nontrivial -- three-loop -- order and discuss some properties of the crumpling-to-flat transition.  

One  reparametrizes the action (\ref{S}) in terms of  $D$ auxiliary $d$-components fields $\{\bm A_{\alpha}\}$, $\alpha=1\dots D$ so that the partition function of the theory reads (using the notation $\bm\varphi$ instead of $\bm R$):
\begin{flalign}
Z=\int {\cal D}\bm \varphi \prod_{\alpha=1}^D {\cal D}\bm A_{\alpha} \ \delta(\bm A_{\alpha}-\partial_{\alpha}\bm\varphi)\  e^{\displaystyle -S[\{\bm A_{\alpha}\}]}\ . 
\nonumber
\end{flalign}
The delta constraint can be raised with the help of a second set of  $D$ auxiliary  $d$-components  fields {$\{\bm B_{\beta}\}$:  
\begin{align}
\begin{split}
    Z=\displaystyle\int {\cal D}\bm \varphi & \hspace{-0.1cm} \displaystyle\prod_{\alpha,\beta=1}^D {\cal D}\bm A_{\alpha} {\cal D}\bm B_{\beta} \  e^{\displaystyle -S[\{\bm A_{\alpha}\}]}  \\
    &\ \times e^{\displaystyle  -\color{black}i  \int d^Dx\,  \bm B_{\alpha}.(\bm A_{\alpha}-\partial_{\alpha} \bm \varphi)} \, . \label{partition2}
\end{split}
\end{align}
In the partition function~(\ref{partition2}), the fields ~$\{\bm B_{\alpha}\}$ and~$\{\bm \varphi\}$ appear only in terms quadratic in the fields. As a consequence, there is no interaction vertex with~$\bm B$ or~$\bm \varphi$ legs. We conclude that there are no 1PI Feynman diagrams with such fields as external legs and the~$\bm B - \bm\varphi$  sector renormalizes trivially; only the auxiliary fields~$\{\bm A_{\alpha}\}$ renormalize nontrivially. 

The matrix of second derivative of  $S'=S+ i \int d^Dx\,  \bm B_{\alpha}.(\bm A_{\alpha}-\partial_{\alpha} \bm \varphi$) in the basis $\{\bm X_{\alpha}\}=(\{\bm A_{\alpha}\}, \{\bm B_{\alpha}\}, \bm\varphi)$  is given, in Fourier space, by: 
\begin{equation}
{\delta^2 S'\over \delta  X_{\alpha}^i  \delta  Y_{\beta}^j}=\begin{pmatrix}
p_{\alpha}p_{\beta}+\delta_{\alpha\beta}\,  r & i \delta_{\alpha\beta} & 0 \\
i \delta_{\alpha\beta} & 0 & - p_{\alpha}\\
0 & p_{\beta} & 0
\end{pmatrix}
\delta_{ij} \, ,
\nonumber
\end{equation}
with the indices $i,j$ running from 1 to $d$.  The inverse matrix provides the propagator
\begin{equation}
G^{(2)\, ij}_{\alpha\beta}=
\begin{pmatrix}
{P^{\parallel}_{\alpha\beta}\over {\bm p}^2+r} & -  i P^{\perp}_{\alpha\beta} & {- i p_{\alpha}\over {\bm p}^2({\bm p}^2+r)} \\
\\
- i P^{\perp}_{\alpha\beta} & r P^{\perp}_{\alpha\beta} & {p_{\alpha}\over {\bm p}^2} \\
\\
i  p_{\beta}\over {\bm p}^2({\bm p}^2+r) & -{p_{\beta}\over {\bm p}^2}  & {1\over {\bm p}^2({\bm p}^2+r)}
\end{pmatrix}
\delta_{ij} \, ,
\nonumber
\end{equation}
where one remarks  that  the propagator of the $\{\bm A_{\alpha}\}$-fields is given by ${P^{\parallel}_{\alpha\beta}/{(\bm p}^2+r)}$. The existence of a longitudinal  propagator is the  only  change  with respect to the standard $\bm\varphi^4$ theory. This is a tremendous simplification compared to the usual, brute force, treatment of derivative field theories.

\begin{center} 
{\bf VI.  RENORMALIZATION  GROUP EQUATION AT TWO LOOP ORDER}
\end{center}

We have derived the two-loop order RG equations for the model (\ref{partition2}) within the modified minimal subtraction $\overline{\hbox{MS}}$ scheme. The   diagrammatic  is the one  of the $\bm\varphi^4$ theory while the vertices  involve two coupling constants $\lambda$ and $\mu$ and a tensorial algebra  associated with  both the vectorial -- Roman -- and derivative -- Greek -- indices. We have used {\it Mathematica} to perform the algebra and  compute the integrals.  One introduces  the renormalized field  $\bm A_{\alpha R}$   through
$\bm A_{\alpha}= Z^{1/2} \bm A_{\alpha R}$ as well  the dimensionless renormalized coupling constants $\lambda_R$  and $\mu_R$ through  $\lambda=k^{\epsilon} Z^{-2}  Z_{\lambda} \lambda_R$ and $\mu=k^{\epsilon} Z^{-2} Z_{\mu} \mu_R$. Here, $k$  is the
running  momentum scale and $\epsilon=4-D$.  Finally, within the $\overline{\hbox{MS}}$ scheme,  one introduces the scale $\bar k^2=4\pi e^{-\gamma_E} k^2$ where $\gamma_E$ is the Euler constant.  One defines the RG flow of the renormalized coupling constants at fixed bare theory $\beta_{\lambda}=\partial_t \lambda$,  $\beta_{\mu}=\partial_t \mu$  and $\beta_r=\partial_t r$ with $t=\log \bar k$ 
where, for simplicity, and from now on,  we omit the index $R$ for the renormalized quantities. The running field anomalous dimension is given by $\eta=-\partial_t \log Z$.  The RG functions at two-loop order are thus  given by:
\begin{align}
    &\beta_{\lambda}(\lambda,\mu) =-\epsilon \lambda \nonumber  \\
    &+c_1\big((6 d+7) \lambda ^2  +2 (3 d+17) \lambda  \mu +(d+15) \mu^2\big) \nonumber \\
    &- \displaystyle{c_1^2\over 6} \big((69 d+52) \lambda^3+(54 d^2-16 d+541)\lambda^2 \mu  \label{betalam} \\
    & +(36 d^2+281 d-110)\lambda\mu ^2+(6 d^2+112 d-95)\mu^3\big) \nonumber \\
\nonumber \\
    & \beta_{\mu}(\lambda,\mu) =-\epsilon \mu+c_1\big(\lambda ^2+ (d+21) \mu ^2+10 \lambda  \mu \big) \nonumber\\
    & +\displaystyle{c_1^2\over 12} \big((96 d+55) \lambda^3+ (470 d+289)\lambda^2 \mu \label{betamu}\\
    & +(146 d +421)\lambda\mu^2+(-212d+475)\mu^3\big) \nonumber \\
\nonumber \\
    &\beta_{r}(r,\lambda,\mu) =- 2r +  3 c_1 r \big((2d+1)\lambda+(d+5) \mu\big) \nonumber \\ 
    &-\displaystyle{3 c_1^2 r\over 4} \big((19d-1) \lambda^2+2(6 d^2+5d+25)\lambda\mu \label{betar}\\ 
    &+(4d^2+41d+27)\mu^2\big) \nonumber
\end{align}
with $c_1=\displaystyle{1/96 \pi ^2}$. 
 This last equation provides the {\sl running}  exponent  $\nu(\lambda,\mu)$: 
\begin{align}
 & \nu(\lambda,\mu) =\displaystyle{1\over 2}+ \displaystyle{3 c_1\over{4}}\Big((2d+1)\lambda +(d+5)\mu\Big) \nonumber \\
 & + \displaystyle{3 c_1^2\over 16}\Big((24 d^2+5d+7) \lambda^2 +2(6 d^2+61d+5) \lambda \mu \nonumber \\
 &+(2 d^2+19d+123) \mu^2\Big) \label{nu}
\end{align}
 while the {\sl running}  field renormalization $\eta(\lambda,\mu)$ at the first nontrivial -- three-loop -- order is given by: 
\begin{align}
\eta(\lambda,\mu)=& \displaystyle {(d+2)(\lambda + 2 \mu) \over 3 (32\pi ^2)^3} \times \label{eta}\\
& \big((2 d+ 3) \lambda^2 + 2 (d + 9) \lambda \mu  +(d+19) \mu^2 \big) \nonumber
\end{align}
Equations (\ref{betalam})-(\ref{eta})  constitute our main results. They generalize to the next nontrivial order the expressions derived by Paczuski  {\it et al.} \cite{paczuski88}. Equations (\ref{betalam}),  (\ref{betamu})  and (\ref{eta})  also generalize to $d\ge 1$-component vector fields  those derived by Safari {\it et al.} \cite{safari22} in the case of  scalar  fields.  To make contact with their expressions, one has to take the limit $d\to 1$  in our  RG functions (\ref{betalam}),  (\ref{betamu}) and (\ref{eta}) and to form the combination $g=\lambda/2+\mu$  whose RG flow  gives that found in  \cite{safari22} while one finds $\eta=5 g^3/(4\pi)^6$ also in agreement with  \cite{safari22}. 

\vspace{0.5cm}

\begin{center} 
{\bf VII.  CRUMPLED-TO-FLAT TRANSITION IN POLYMERIZED MEMBRANES}
\end{center}

As said, the  model (\ref{S}) has been investigated perturbatively at one-loop order within an $\epsilon$-expansion in \cite{paczuski88} and by other techniques  \cite{david88,aronovitz89,ledoussal92,ledoussal18,kownacki09,braghin10,hasselmann11,essafi11,essafi14,coquand18}.  In the vicinity of $D = 4$, the one-loop computation has allowed to identify a phenomenon of fluctuation-induced first order phase transition: below a critical value of the dimension $d$, approximately equal to $d_c\sim  218$, no stable fixed point is found so that the RG flow runs away at infinity and the transition is expected to be of first order, while above $d_c$ one finds a nontrivial fixed point and the transition is expected to be of second order.  A large amount of numerical simulations have been performed on this model in order to determine the nature of the phase transition in the physical -- $D=2, d=3$ -- case, see e.g. the contributions  of Cantor and  of  Gomper and  Kroll  in \cite{proceedings89}.  These results have  been controversial as they have led   to conclude  either to  a first order transition \cite{kownacki02,koibuchi04} or to a second order one  \cite{cuerno16}.  The nonperturbative approaches, performed also directly in $D=2$ and $d=3$ dimensions, have also led to various kinds of transitions, second \cite{kownacki09}  or  first order  \cite{essafi11}, according to the kind of truncation used. 

\bigskip 
{\it One-loop order.}  Let us now discuss  our predictions for the crumpled-to-flat transition. One starts  with the one-loop order results that have not  been explicitly given  in the past literature.  At leading order in $\epsilon$, one finds four solutions in agreement with \cite{paczuski88}. They are given by: $\lambda={(16\pi^2/3)}\Omega\,   \epsilon$ and $\mu={(16\pi^2/ 3)}X[\Omega] \Omega\,  \epsilon$ with 
$\Omega=0, \Omega=\Omega[1,1,d], \Omega=\Omega[x_0,y_0,d]$ and $\Omega=\Omega[y_0,x_0,d]$ where the function $\Omega[x,y,d]$, as well as the parameters $x_0$ and $y_0$, are given in the Appendix.  One considers the case $D<4$. The  value $\Omega=0$  corresponds to  the Gaussian fixed point which is always twice unstable \footnote{One omits the always relevant direction associated with the temperature.} while the value $\Omega[y_0,x_0,d]$ corresponds to a once unstable one. The stability of the other fixed points depends on the value of $d$ with respect to a critical value $d_c$. For $d>d_c$, the  value $\Omega[1,1,d]$ corresponds to a once unstable fixed  point and  the value $\Omega[x_0,y_0,d]$  to a stable one. This last fixed point  controls the second order, crumpled-to-flat, transition. For  $d<d_c$, the fixed points associated with  the values $\Omega[1,1,d]$ and  $\Omega[x_0,y_0,d]$   are complex  ones with conjugate coordinates.  There is no longer any  fully stable fixed point, and the transition is expected to be of first order. At one-loop order, the critical value $d_c$ is given by the  root of a  polynomial, see the Appendix,  and is given by $d_c\simeq 218.20$. 

\bigskip 
{\it Two-loop order.}  At two-loop order, the situation is not qualitatively modified. There are still four fixed points and, for $d>d_c$, only one,  corresponding to  $\Omega[x_0,y_0,d]$,  is  fully stable. The coordinates of the fixed point are given in the Appendix.  With the help of these coordinates, one can compute several physical quantities at second order in $\epsilon$ like the critical exponents $\nu$ and $\eta$. However, their full $d$-dependence is extremely involved and not very useful to explicit. Also, as seen from the value of $d_c$,  in the vicinity of the upper critical dimension $D=4$, the second order phase transition occurs at large values of the embedding dimension $d$. It is thus  relevant to evaluate the critical quantities within a $1/d$ expansion. From the expression of $\nu(\lambda,\mu)$, Eq.~(\ref{nu}), one finds at orders $1/d$ and  $\epsilon^2$ at the stable fixed point: 
\begin{equation}
\nu={1\over 2} + \left({1\over 4} - {33\over 2 d}\right) \epsilon+ \left({1\over 8} + {129\over 8 d}\right)\epsilon^2 + {\cal{O}}\left({\epsilon\over d^2},{\epsilon^2\over d^2},\epsilon^3\right)\ 
\label{nud}
\end{equation}
that coincides exactly with the results following the  $1/d$ analysis of Paczuski and Kardar~\cite{paczuski89} when it is further expanded in powers of $\epsilon$.  In the same way, from the expression  of  $\eta$, Eq.~(\ref{eta}), one finds, at the stable fixed point:
\begin{equation}
\eta={25 \over 3 d}  \epsilon^3+ {\cal{O}}\left({\epsilon^3\over d^2},\epsilon^4\right)\ 
\label{etad} 
\end{equation}
that also coincides  with  the expression obtained in~\cite{paczuski89}. Expressions~(\ref{nud}) and~(\ref{etad}) provide strong checks of our computations, while higher order in $1/d$ or even the full expressions can be easily derived from  the Appendix. A quantity of strong interest  is the critical value $d_c$ at this order. It is computed  by requiring that the coordinates of the stable fixed point develop an  imaginary part when $d\to d_c^-$. One finds :
\begin{equation}
d_c(\epsilon)=218.20-448.25\,  \epsilon + {\cal{O}}(\epsilon^2)
\label{nc}
\end{equation}
where we  emphasize that the correction of order $\epsilon$ would have been extremely painful to get without  the present  auxiliary  fields formalism. 

As seen on expression~(\ref{nc}), the correction of order $\epsilon$ is large and of the same order of magnitude than the dominant term. Note that this is a very generic feature of  fluctuation-induced  first order  phase transitions, where there exists a critical number $N_c$ of the  number of components $N$ of the order parameter above which the transition is of second order and under which it is of first order.  One has, for instance,  in frustrated magnets, see e.g. \cite{delamotte04} $N_c(\epsilon)=21.8-23.4\, \epsilon$ or in electroweak phase transition \cite{arnold94} $N_c(\epsilon)=718-990.83   \, \epsilon$. 
In order to get trustable predictions as for the nature of the crumpled-to-flat phase transition in the physical dimension $D = 2$, several techniques can be used. One can -- easily with the present, auxiliary  fields, approach  --  push  the $\epsilon$-expansion  at  higher orders  and  resum  the series $d_c(\epsilon)$  obtained. 
However, obtaining a converged result in $D=2$ starting from the upper critical dimension $D=4$ is likely to be an extremely difficult task in practice, requiring to compute high-order terms in the series expansion and to employ  sophisticated resummation techniques.  Still with the help of the auxiliary  fields approach one could, alternatively, make use of the functional RG using high-order truncations in the fields and field-derivatives.  However, investigations  done in \cite{essafi14}, where  a field expansion up to order $\varphi^{8}$  has been employed,  have revealed strong variations of the results with the order of the truncation. This again implies that high-order truncations or even the full field content may be required to obtain converged results. Finally, conformal bootstrap techniques, that have allowed to get well-converged exponents in the context of the standard $\varphi^4$ theory, are  very  promising  provided that  the  conformal invariance of action~\eqref{S} at the fixed point is definitely established and the techniques usable.

 \bigskip 
 
 \begin{center} 
{\bf VIII. CONCLUSION}
\end{center}

 We have introduced  a technique allowing to investigate and determine the RG properties of derivative shift theories with the same level of technicality as  usual -- nonderivative --  theories. This allowed us  to derive the two-loop RG equations of the $\bm\varphi^4$ derivative theory, also relevant to study the crumpled-to-flat transition in membranes thirty-five years after the one-loop  computation of Paczuski {\it et al.}. This technique  can be used to extend our computation at higher orders. It can also be easily generalized to a situation  involving  several fields, as in the  flat phase of membranes, see action  (\ref{action}), where phonons and flexural degrees of freedom coexist, even if this case has already been studied at high (four) orders within a loop expansion.   Still in the context of membranes,  our  approach  is also relevant to investigate   fluid membranes where one has  moreover  to deal with  gauge -- diffeomorphism -- invariance \cite{proceedings89,codello11}. Finally, our  technique  can be of great interest and can be easily implemented in various  contexts around that of quantum gravity, as Galileon theory, Hor$\check{\hbox{a}}$va-Lifshitz gravity, string, brane theory,~$\dots$ either in a perturbative or in a nonperturbative framework.

\acknowledgements

L.D. and D.M.  greatly thank S. Metayer for his contribution in early investigations of the model studied  here by means of a usual perturbative approach. C. D., D.M. and  M.T. thank R. Ben Al\`i Zinati for discussions. 

\begin{widetext}

\begin{center} 
{\bf APPENDIX: FIXED POINTS AT ONE AND TWO-LOOP ORDER} 
\end{center}

At one-loop order, the four fixed points are given by: 
\begin{equation}
\lambda_1=\frac{16\pi^2}{3}\Omega\,   \epsilon~~~~~\textrm{and}~~~~~\mu_1=\frac{16\pi^2}{3}X[\Omega] \Omega\,  \epsilon
\end{equation}
with $\Omega=0, ~\Omega=\Omega[1,1,d], ~\Omega=\Omega[x_0,y_0,d]$ and $\Omega=\Omega[y_0,x_0,d]$ and the parameters $x_0=(-1+i\sqrt{3})/2$ and $y_0=-(1+i\sqrt{3})/2
$. The functions $\Omega[x,y,d]$ and $X[\Omega]$ are defined by:
\begin{equation}
\begin{aligned}
&\Omega[x,y,d]=\frac{1}{P_1(d)}\left(4 P_2(d)+x \frac{P_3(d)}{T^{1/3}}+ y\, T^{1/3}\right)\hspace{0.7cm} {\hbox{and}} \hspace{0.7cm} X[\Omega]=\frac{(3d +63 -P_4(d)\,  \Omega)}{ (3d +45 + P_5(d)\,  \Omega)}
\end{aligned}
\end{equation}
where one defines $T=P_6(d)+3\sqrt{3} \sqrt{\Delta}$ and $\Delta=-P_1(d)^2  P_7(d)^2 P_8(d)$
with $P_1(d),\dots,P_8(d)$ given by: 
\begin{equation}
\begin{aligned}
P_1(d)= &80 + 28 d - 297 d^2 - 35 d^3 - d^4\\
P_2(d)= &146 - 30 d + 18 d^2 + d^3\\
P_3(d)=&372736 - 129312 d - 18960 d^2 - 25493 d^3 + 3042 d^4 + 474 d^5 + 13 d^6\\
P_4(d)= &22 + 22 d + d^2 \\
P_5(d)= &94 + 25 d + d^2\\
P_6(d)= &219152384 - 124941312 d + 53895600 d^2 + 14104092 d^3 - 103622787 d^4 - 34822215 d^5 -\\
&3884853 d^6 - 193509 d^7 - 4365 d^8 - 35 d^9\\
P_7(d)= &2304+971d+83 d^2+2d^3   \\  
P_8(d)= &4096-5376d+3765 d^2-1981d^3+9d^4    
\end{aligned}
\end{equation}

For the fixed point parametrized by $\Omega[x_0,y_0,d]$, the critical dimension $d_c$ above which the transition is of second order  is determined by finding the value of $d$ that cancels the imaginary part of the fixed point. Here, it can be seen that it is directly determined by the sign of $\Delta$, hence the root of the polynomial $P_8(d)$. One obtains $d_c\simeq 218.20$.\\

At two-loop order, the four fixed points are given by:
\begin{equation}
\begin{aligned} 
\lambda_2=&\frac{1}{A(d)}\left[(485+829d+96d^2)\lambda_1^4 + (4444+4398d+728d^2)\lambda_1^3\mu_1+(11038+4768d+1168d^2+36d^3) \lambda_1^2\mu_1^2+\right.\\
&(2940+4378d+548d^2+24d^3) \lambda_1 \mu_1^3+(1045+603d+88d^2+4d^3)\mu_1^4 -\\
   &32 \pi^2 \left((52+69d)\lambda_1^3+(541-16d+54d^2)\lambda_1^2\mu_1-(110-281d-36d^2)\lambda_1\mu_1^2-(95-112d-6d^2) \mu_1^3\right)\left.\right]\\
\mu_2=&\frac{-1}{A(d)}\left[(163 + 380d+192d^2)\lambda_1^4 + (1520 + 2493d + 1072d^2)\lambda_1^3 \mu_1 + (4350 + 4269d + 966d^2)\lambda_1^2\mu_1^2 + \right.\\
&(3064 + 2715d - 154d^2)\lambda_1 \mu_1^3 + (2375 - 353d - 192d^2) \mu_1^4 -\\
&16 \pi^2 \left((55 + 96d)\lambda_1^3 + (289 + 470d) \lambda_1^2\mu_1 + (421 + 146d) \lambda_1 \mu_1^2 + (475 - 212d) \mu_1^3\right)\left.\right]
\end{aligned}
\end{equation}
where $A(d)$ is defined by:
\begin{equation}
A(d) = 2304\pi^2 (768 \pi^4 + (6 + 9d)\lambda_1^2 + 2 P_4(d) \lambda_1 \mu_1 + P_5(d) \mu_1^2 - 32 \pi^2 ((6 + 3~d) \lambda_1 + (19 + 2~d) \mu_1)).
\end{equation}
Using these coordinates, the critical value of $d_c$ at the $\epsilon$-order can be computed. This is done by expanding $d_c$ into $d_{c0}+\epsilon~d_{c1}$ and requiring that $d_{c1}$ takes the value under which the coordinates of the stable fixed point develop an imaginary part. One finds:
\begin{equation}
d_c(\epsilon)=218.20-448.25\,  \epsilon + {\cal{O}}(\epsilon^2)\ 
\label{nc2}
\end{equation}

\end{widetext}


\end{document}